\documentclass[aps,preprint]{revtex4}%
\usepackage{amsfonts}
\usepackage{amsmath}
\usepackage{amssymb}
\usepackage{graphicx}%
\begin{document}
\preprint{ }
\title[Short title for running header]{Cooperative spin decoherence and population transfer}
\author{C. Genes}
\affiliation{Michigan Center for Theoretical Physics, FOCUS Center, and Physics Department,
University of Michigan, Ann Arbor 48109-1040, USA}
\author{P. R. Berman}
\affiliation{Michigan Center for Theoretical Physics, FOCUS Center, and Physics Department,
University of Michigan, Ann Arbor 48109-1040, USA}
\pacs{PACS number}

\begin{abstract}
An ensemble of multilevel atoms is a good candidate for a quantum information
storage device. The information is encrypted in the collective ground state
atomic coherence, which, in the absence of external excitation, is decoupled
from the vacuum and therefore decoherence free. However, in the process of
manipulation of atoms with light pulses (writing, reading), one inadvertently
introduces a coupling to the environment, i.e. a source of decoherence. The
dissipation process is often treated as an independent process for each atom
in the ensemble, an approach which fails at large atomic optical depths where
cooperative effects must be taken into account. In this paper, the cooperative
behavior of spin decoherence and population transfer for a system of two,
driven multilevel-atoms is studied. Not surprisingly, an enhancement in the
decoherence rate is found, when the atoms are separated by a distance that is
small compared to an optical wavelength; however, it is found that this rate
increases even further for somewhat larger separations for atoms aligned along
the direction of the driving field's propagation vector. A treatment of the
cooperative modification of optical pumping rates and an effect of
polarization swapping between atoms is also discussed, lending additional
insight into the origin of the collective decay.

\end{abstract}
\volumeyear{year}
\volumenumber{number}
\issuenumber{number}
\eid{identifier}
\date[Date text]{date}
\received[Received text]{date}

\revised[Revised text]{date}

\accepted[Accepted text]{date}

\published[Published text]{date}

\startpage{1}
\endpage{ }
\maketitle

\section{Introduction\bigskip}

A system of atoms interacting with a common reservoir (the electromagnetic
vacuum) is often treated using an assumption of independent dissipation rates
for the atoms. This is a valid assumption for the case when the interatomic
distances are large (compared to an optical wavelength); however, when the
distances between atoms become smaller or comparable to an optical wavelength,
the mode structure around one atom is changed due to the presence of other
atoms located in its immediate vicinity, and the decay rates are modified.
Equivalently,\ the radiation emitted by one atom can be scattered off the
second atom, thus changing the radiative properties of the system. As a
consequence, the radiative decay of the ensemble must be viewed as a
cooperative effect. A quantitative analysis of cooperative effects was given
by Dicke \cite{dicke} for an ensemble of two-level systems confined to a
spherical volume whose radius is much smaller than an optical wavelength. For
such a system, the collective decay rate can be increased by a factor
proportional to the number of atoms in the ensemble. Further investigations
extend the treatment to arbitrary interatomic distances, although the
calculations become more complex. Formalisms for the treatment of cooperative
spontaneous emission and resonance fluorescence from a system of many atoms
have been developed \cite{ernst-stehle, lehmberg1, agarwal, eberly, walls},
and have been applied to a system of two two-level atoms \cite{lehmberg2,
ficek, richter}.

Cooperative decay in multilevel atomic systems has yet to be treated in
detail. A multilevel atom has properties not possessed by a two-level atom:
for example, it can store information in superpositions of ground state
magnetic sublevels. Such ensembles are extensively discussed in the literature
as convenient systems for information storage or large scale entanglement
generation \cite{entanglement}. In particular, pencil-shaped media have been
used for the generation of spin squeezed and Schrodinger cat states, in the
context of continuous measurement of a scattered field
\cite{pencil-entanglement}. In such schemes, the coupling to the vacuum has a
two-fold function: on one hand it gives rise to the signal while, on the other
hand, it leads to an irreversible leakage of information from the system to
the environment. In treating the losses due to spontaneous emission, the above
mentioned assumption of independent atoms is generally used, which is a sound
assumption as long as the atomic density is low. However, optimal results
(e.g. strong entanglement) are found in the regime of resonant optical depths
greater than unity, a regime in which pencil-shaped media of two-level atoms
exhibit superradiant behavior \cite{pencil}. Even if the dynamics of a
collection of multilevel atoms might be substantially different from that of
the two-level ensemble, the validity of the independent spontaneous emission
regime is questionable.

We proceed in the present publication with an analysis of cooperative effects
in a system of two, four-level atoms. This provides the starting point for an
extension to many atom systems, while it also addresses the non-trivial
question of the importance of cooperative decoherence in a simple quantum
information system of two qubits. The calculations are performed for a
$J=1/2\rightarrow J^{\prime}=1/2$ transition irradiated with a monochromatic,
off-resonant, $\sigma_{+}$ polarized, classical laser field. The collective
decoherence of an initial equal superposition of ground sublevels ($x$
polarized atomic state) is obtained for arbitrary interatomic separations and
compared with that of atoms independently coupled to the reservoir. In
addition, the transfer of population from a $z$ polarized atomic state with
both atoms in one of the ground sublevels to another $z$ polarized state with
both atoms in the other sublevel, is analyzed. A polarization swap effect is
also discussed, where an $x$ polarized atom induces $x$ coherence in a
neighboring atom \cite{footnote}.

The paper is organized as follows: in Sec. II the theoretical method used is
described. In Sec. III analytical solutions for the case of close atoms are
obtained. Numerical solutions for arbitrary separations are discussed and
plotted in Sec. IV. In Sec. V the polarization swap effect for arbitrary
separations is discussed, while Sec. VI contains some conclusions.

\section{Theory}

\begin{figure}
[h]
\begin{center}
\includegraphics[
height=3.4714in,
width=2.3419in
]%
{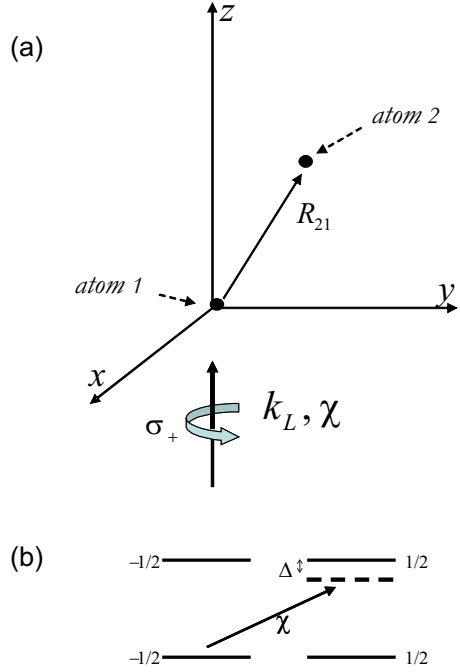}%
\caption{{\footnotesize (a) The classical field is propagating along the }$z$
{\footnotesize axis, driving the two atoms, located at the origin }$R_{1}=0$
{\footnotesize and at }$R_{2}=R_{21}${\footnotesize , respectively. (b)
Internal structure of a single atom. \ \ \ \ }}%
\label{scheme}%
\end{center}
\end{figure}

As indicated in Fig. \ref{scheme}(a), the two atoms (having natural frequency
$\omega_{0}$) are located at the origin, $\mathbf{R}_{1}=0$, and at
$\mathbf{R}_{2}=\mathbf{R}_{21}$, respectively. The traveling wave driving
field propagates in the positive $z$ direction with wave vector $\mathbf{k}$,
frequency $\Omega$ (detuned from $\omega_{0}~$by $\Delta$), and circular
polarization $\sigma_{+}$.%
%TCIMACRO{\FRAME{fhFU}{2.3419in}{3.4714in}{0pt}{\Qcb{{\footnotesize (a) The
%classical field is propagating along the }$z$ {\footnotesize axis, driving the
%two atoms, located at the origin }$R_{1}=0$ {\footnotesize and at }%
%$R_{2}=R_{21}${\footnotesize , respectively. (b) Internal structure of a
%single atom. \ \ \ \ }}}{\Qlb{scheme}}{Fig1.eps}%
%width 2.3419in;  height 3.4714in;  depth 0pt;  original-width 6.4212in;
%original-height 9.5441in;  cropleft "0";  croptop "1";  cropright "1";
%cropbottom "0";  filename 'Fig1.eps';file-properties "XNPEU";}}}%
%BeginExpansion
%EndExpansion
Denoting the $4$ states of a single atom with $\left\vert \downarrow
\right\rangle $ (ground, $m_{\downarrow~}=-1/2$ eigenvalue), $\left\vert
\uparrow\right\rangle $ (ground, $m_{\uparrow~}=1/2$ eigenvalue), $\left\vert
\beta\right\rangle $ (excited, $m_{\beta~}=-1/2$ eigenvalue) and $\left\vert
\alpha\right\rangle $ (excited, $m_{\alpha~}=1/2$ eigenvalue), the classical
field drives the $\left\vert \downarrow\right\rangle \rightarrow$ $\left\vert
\alpha\right\rangle $ transitions. The Hamiltonian for the system (composed of
the two atoms indexed with $i$, where $i=1,2$) is a sum of the free
Hamiltonian ($H_{0}$), the classical field-atom interaction Hamiltonian
($V_{c}$) and quantized bath-atom interaction Hamiltonian ($V_{q}$) given by:%
\begin{align}
H_{0~}  &  =\underset{i=1,2}{%
%TCIMACRO{\tsum }%
%BeginExpansion
{\textstyle\sum}
%EndExpansion
}\hbar\omega_{0}\left[  \left\vert \alpha\right\rangle _{i}\left\langle
\alpha\right\vert _{i~}+\left\vert \beta\right\rangle _{i}\left\langle
\beta\right\vert _{i}\right]  +\underset{\mathbf{k},\lambda}{%
%TCIMACRO{\tsum }%
%BeginExpansion
{\textstyle\sum}
%EndExpansion
}\hbar\omega_{k}a_{\mathbf{k},\lambda}^{\dagger}a_{\mathbf{k},\lambda~},\\
V_{c~}  &  =\underset{i=1,2}{%
%TCIMACRO{\tsum }%
%BeginExpansion
{\textstyle\sum}
%EndExpansion
}\hbar\chi\left[  \left\vert \alpha\right\rangle _{i}\left\langle
\downarrow\right\vert _{i~}e^{i\mathbf{k}_{L}\mathbf{\cdot R}_{i~}}e^{-i\Omega
t}+\left\vert \downarrow\right\rangle _{i}\left\langle \alpha\right\vert
_{i~}e^{-i\mathbf{k}_{L}\mathbf{\cdot R}_{i~}}e^{i\Omega t}\right]
,\nonumber\\
V_{q~}  &  =\underset{i=1,2}{%
%TCIMACRO{\tsum }%
%BeginExpansion
{\textstyle\sum}
%EndExpansion
}\left[  \underset{j=\downarrow,\uparrow}{%
%TCIMACRO{\tsum }%
%BeginExpansion
{\textstyle\sum}
%EndExpansion
}\underset{j^{\prime}=\beta,\alpha}{%
%TCIMACRO{\tsum }%
%BeginExpansion
{\textstyle\sum}
%EndExpansion
}\underset{\mathbf{k},\lambda}{%
%TCIMACRO{\tsum }%
%BeginExpansion
{\textstyle\sum}
%EndExpansion
}\hbar\left[  g_{\mathbf{k},\lambda~}^{jj^{\prime}}\left\vert j\right\rangle
_{i}\left\langle j^{\prime}\right\vert _{i~}a_{\mathbf{k},\lambda~}^{\dagger
}e^{-i\mathbf{k\cdot R}_{i~}}+\left(  g_{\mathbf{k},\lambda~}^{jj^{\prime}%
}\right)  ^{\ast}\left\vert j^{\prime}\right\rangle _{i}\left\langle
j\right\vert _{i~}a_{\mathbf{k},\lambda~}e^{i\mathbf{k\cdot R}_{i~}}\right]
\right]  .\nonumber
\end{align}
As usual, the field is quantized in a volume $\mathcal{V}$ and is described by
the creation and annihilation operators $a_{\mathbf{k},\lambda~}^{\dagger}$and
$a_{\mathbf{k},\lambda}$, corresponding to modes with wave vector $\mathbf{k}$
and polarization $\lambda$. The atomic dipole moment $\mathbf{d}$ (the same
for both atoms) couples to both the classical and quantized fields. The cw
driving field at the position of the atoms ($\mathbf{R}_{i~}$ for $i=1,2$) is
expressed as $\mathbf{E}\left(  \mathbf{R},t\right)  \mathbf{=}\frac{1}%
{2}\left[  E_{0~}e^{i\mathbf{k}_{L~}\mathbf{\cdot R}_{i~}}e^{-i\Omega
t}+cc\right]  \widehat{\mathbf{\epsilon}}_{+~}$. The classical part of the
interaction contains the Rabi frequency defined as:%
\begin{equation}
\chi=\frac{d_{+}E_{0}}{2\hbar},
\end{equation}
where $d_{+}$ is the matrix element $d_{+}=\left\langle \alpha\right\vert
\mathbf{d}\cdot\widehat{\mathbf{\epsilon}}_{+}\left\vert \downarrow
\right\rangle $; the interaction with the quantum vacuum has an associated
coupling strength%
\begin{equation}
g_{\mathbf{k},\lambda~}^{jj^{\prime}}=-i\left(  \frac{\omega_{k~}}%
{2\epsilon_{0}\hbar\mathcal{V}}\right)  ^{1/2}d_{\mathbf{k},\lambda
~}^{jj^{\prime}}%
\end{equation}
proportional to the dipole matrix element in the direction of the unit
polarization vector $d_{\mathbf{k},\lambda~}^{jj^{\prime}}=\left\langle
j\right\vert \mathbf{d}\cdot\widehat{\mathbf{\epsilon}}_{\mathbf{k},\lambda
~}\left\vert j^{\prime}\right\rangle $.

The quantities that are relevant in what follows are the collective coherence
operator%
\begin{equation}
P_{\downarrow\uparrow~}=\underset{i=1,2}{%
%TCIMACRO{\tsum }%
%BeginExpansion
{\textstyle\sum}
%EndExpansion
}\left\vert \downarrow\right\rangle _{i}\left\langle \uparrow\right\vert _{i~}%
\end{equation}
and the normalized population operator%
\begin{equation}
P_{\uparrow\uparrow~}=\frac{1}{2}\underset{i=1,2}{%
%TCIMACRO{\tsum }%
%BeginExpansion
{\textstyle\sum}
%EndExpansion
}\left\vert \uparrow\right\rangle _{i}\left\langle \uparrow\right\vert _{i~}.
\end{equation}
The derivation of the time evolution of the expectation values of these two
operators is the goal of our calculations. They can expressed in terms of
density matrix elements as%
\begin{align}
\left\langle P_{\downarrow\uparrow~}\right\rangle  &  =\rho_{\uparrow
\uparrow;\downarrow\uparrow~}+\rho_{\uparrow\downarrow;\downarrow\downarrow
~}+\rho_{\uparrow\uparrow;\uparrow\downarrow~}+\rho_{\downarrow\uparrow
;\downarrow\downarrow~},\label{cohandpop}\\
\left\langle P_{\uparrow\uparrow~}\right\rangle  &  =\frac{1}{2}\left(
2\rho_{\uparrow\uparrow;\uparrow\uparrow~}+\rho_{\uparrow\downarrow
;\uparrow\downarrow~}+\rho_{\downarrow\uparrow;\downarrow\uparrow~}\right)
.\nonumber
\end{align}
$~$\ We consider two problems: first, one with both atoms prepared initially
in a superposition of ground states with maximum coherence $\frac{1}{\sqrt{2}%
}\left(  \left\vert \downarrow\right\rangle +\left\vert \uparrow\right\rangle
\right)  $ (collective coherence equal to $1$) and, second, one where the
population is transferred from the state with both atoms in the $\left\vert
\downarrow\right\rangle $ state to the one with both atoms in the $\left\vert
\uparrow\right\rangle $ state.

Qualitative (and some quantitative) details of the calculations are described
in the following. The subspace of interest in which the collective operators
defined above act (henceforth named the ground subspace) is of dimension $4$
and it is spanned by state vectors containing the ground substates of the two
atoms ( $\left\vert \downarrow\downarrow\right\rangle $,$\left\vert
\downarrow\uparrow\right\rangle $, $\left\vert \uparrow\downarrow\right\rangle
$ and $\left\vert \uparrow\uparrow\right\rangle $). A set of $16$ density
matrix equations completely describes the dynamics of this space. However, the
ground subspace is coupled to the ground-excited subspace of dimension $8$
(containing states with one excitation as for example $\left\vert
\downarrow\alpha\right\rangle $) through the classical field. This is, in its
turn, coupled to the excited subspace of dimension $4$ (containing states of
two excitations like $\left\vert \alpha\alpha\right\rangle $) which can decay
back to the ground states. In a density operator approach, a total of $256$
density matrix elements coupled to each other come into play, which makes the
task at hand extremely complex.

Some simplifications are possible. First, terms occurring in the evolution of
the ground state density matrix elements are separated into in-terms (due to
spontaneous emission from upper states) and out-terms (driving terms due to
the presence of the classical field), and these terms are treated separately.
Second, an amplitude rather than a density matrix approach is sufficient to
obtain expressions for the excited-ground and excited subspace density matrix
elements which enter the equations. The procedure is described in Appendix B
where it is applied to the derivation of the decoherence and population
transfer rate for a single 4-level atom. The treatment is perturbative in the
sense that excited state populations are assumed to be negligibly small, as in
many treatments of optical pumping.

The states coupled by the fields in this approximation are denoted by
$\left\vert \mu^{\prime}\nu\right\rangle $, $\left\vert \mu\nu^{\prime
}\right\rangle $ and $\left\vert \mu\nu\right\rangle $ where the convention
used is that the prime indicates excited states ($\alpha$ or $\beta$), while
the unprimed symbols indicate ground states ($\uparrow$ or $\downarrow$). By
eliminating the intermediate states involving the radiation field (procedure
outlined in Refs. \cite{hao-berman, berman-milonni}), one can obtain the
following coupled equations of motion for the state amplitudes containing one
excitation ($b_{\mu\nu^{\prime}}$ and $b_{\mu^{\prime}\nu}$), in an
interaction picture:%
\begin{align}
\overset{\cdot}{b}_{\mu^{\prime}\nu~}  &  =-\gamma b_{\mu^{\prime}\nu~}%
-\gamma\underset{\mu,\nu^{\prime}}{%
%TCIMACRO{\tsum }%
%BeginExpansion
{\textstyle\sum}
%EndExpansion
}G_{m_{\mu^{\prime}}-m_{\mu~};m_{\nu^{\prime}}-m_{\nu~}}(\mathbf{R}%
_{21})\left\{  m_{\mu~},m_{\mu^{\prime}~}\right\}  \left\{  m_{\nu~}%
,m_{\nu^{\prime}}\right\}  b_{\mu\nu^{\prime}~}\label{AmpEqs}\\
&  -i\Delta b_{\mu^{\prime}\nu~}+i\chi e^{i\mathbf{k}_{L}\mathbf{\cdot R}%
_{1~}}b_{\downarrow\nu~}\delta_{m_{\mu^{\prime}},1/2~},\nonumber\\
\overset{\cdot}{b}_{\mu\nu^{\prime}~}  &  =-\gamma b_{\mu\nu^{\prime}~}%
-\gamma\underset{\mu,\nu^{\prime}}{%
%TCIMACRO{\tsum }%
%BeginExpansion
{\textstyle\sum}
%EndExpansion
}G_{m_{\mu~}-m_{\mu^{\prime}};m_{\nu~}-m_{\nu^{\prime}}}(\mathbf{R}%
_{12})\left\{  m_{\mu^{\prime}~},m_{\mu~}\right\}  \left\{  m_{\nu^{\prime}%
~},m_{\nu~}\right\}  b_{\mu^{\prime}\nu~}\nonumber\\
&  -i\Delta b_{\mu\nu^{\prime}~}+i\chi e^{i\mathbf{k}_{L}\mathbf{\cdot R}%
_{2~}~}b_{\mu\downarrow~}\delta_{1/2,m_{\nu^{\prime}}~}.\nonumber
\end{align}
In the equation above for $\overset{\cdot}{b}_{\mu^{\prime}\nu}$, the first
term on the right-hand side is the decay (at a rate $\gamma$ equal to half the
excited state population decay rate) of the excited state amplitude of an atom
independently coupled to the quantum vacuum. The second term contains a
propagator $G_{m_{\mu^{\prime}}-m_{\mu};m_{\nu^{\prime}}-m_{\nu}}\left(
\mathbf{R}_{21}\right)  $ which includes the effects of the radiation exchange
between atoms: an atom makes a transition from state $\nu^{\prime}$ (quantum
number $m_{\nu^{\prime}}$) to state $\nu$ which is accompanied by a transition
in the other atom ($\mu$ to $\mu^{\prime}$). The real part of $G$ gives a
contribution to the decay rate that varies from $1$, for maximum cooperation
between atoms when their separation is much less than $\lambda$, to $0$, when
no exchange of radiation between atoms is present (infinite separation). The
imaginary part leads to a shift in energy (which adds to $\Delta$ in the
equations above) and varies from $0$ (large separation) to infinity when the
atoms are in the same location. If the minimum interatomic separation is small
but finite, the shift can always be kept small compared to the detuning and
can be neglected. The explicit expressions for the propagators involved in
this problem are given in Appendix A. The geometrical information on the
radiation exchange is contained in the Clebsch-Gordan coefficients $\left\{
m,m^{\prime}\right\}  \equiv\left\langle 1/2,1/2;m,m^{\prime}|1,m-m^{\prime
}\right\rangle $. The last two terms in the right-hand side arise from driving
field induced transitions and from the off-resonant nature of the interaction.

To calculate the out-terms, we note that, as a result of the driving field,
ground state amplitudes ($b_{\mu\nu~}$) are coupled to excited state
amplitudes via%
\begin{equation}
\overset{\cdot}{b}_{\mu\nu~}=i\chi\left[  e^{-i\mathbf{k}_{L}\mathbf{\cdot
R}_{1~}}b_{\alpha\nu~}\delta_{-1/2,m_{\mu~}~}+e^{-i\mathbf{k}_{L}\mathbf{\cdot
R}_{2~}}b_{\mu\alpha~}\delta_{-1/2,m_{\nu~}~}\right]  . \label{gr. st. amp}%
\end{equation}
As a result one finds%
\begin{align}
\overset{\cdot}{\rho}_{\mu\nu;mn}^{out}  &  =\overset{\cdot}{b}_{\mu\nu
~}b_{mn~}^{\ast}+b_{\mu\nu~}\overset{\cdot}{b}_{mn~~}^{\ast}=\left[  i\chi
e^{-i\mathbf{k}_{L}\mathbf{\cdot R}_{1~}}b_{\alpha\nu~}\delta_{-1/2,m_{\mu~}%
~}+e^{-i\mathbf{k}_{L}\mathbf{\cdot R}_{2~}}b_{\mu\alpha~}\delta
_{-1/2,m_{\nu~}~}\right]  b_{mn~}^{\ast}+\label{out-terms}\\
&  +b_{\mu\nu~}\left[  i\chi e^{-i\mathbf{k}_{L}\mathbf{\cdot R}_{1~}%
}b_{\alpha n~}\delta_{-1/2,m_{m~}~}+e^{-i\mathbf{k}_{L}\mathbf{\cdot R}_{2~}%
}b_{m\alpha~}\delta_{-1/2,m_{n~}~}\right]  .\nonumber
\end{align}
The system of equations (Eqs. (\ref{AmpEqs})) is solved for the $8$ amplitudes
$b_{\mu^{\prime}\nu~}$ and $b_{\mu\nu^{\prime}~}$ as functions of the $4$
ground state amplitudes $b_{\mu\nu~}$; these expressions are replaced in the
above equation and with the identification $b_{\mu\nu~}b_{mn~}^{\ast
}\rightarrow\rho_{\mu\nu;mn~}$, the rate equations for $\overset{\cdot}{\rho
}_{\mu\nu;mn}^{out}$ are obtained in terms of the $16$ density matrix elements
$\rho_{\mu\nu;mn~}$(with $\mu,\nu,m,n=\downarrow,\uparrow$).

Next, repopulation from the upper states to the lower states is taken into
account (in-terms)%
\begin{align}
\overset{\cdot}{\rho}_{mn;\mu\nu}^{in}  &  =\underset{m^{\prime},\mu^{\prime}%
}{%
%TCIMACRO{\tsum }%
%BeginExpansion
{\textstyle\sum}
%EndExpansion
}\Gamma_{m\mu~}^{m^{\prime}\mu^{\prime}}\rho_{m^{\prime}n;\mu^{\prime}\nu
~~}+\underset{n^{\prime},\nu^{\prime}}{%
%TCIMACRO{\tsum }%
%BeginExpansion
{\textstyle\sum}
%EndExpansion
}\Gamma_{n\nu~}^{n^{\prime}\nu^{\prime}}\rho_{mn^{\prime};\mu\nu^{\prime}%
~}+\label{in-terms}\\
&  (2\gamma)\underset{m^{\prime},\nu^{\prime}}{%
%TCIMACRO{\tsum }%
%BeginExpansion
{\textstyle\sum}
%EndExpansion
}G_{m_{\nu^{\prime}}-m_{\nu};m_{m^{\prime}}-m_{m}}(\mathbf{R}_{21})\left\{
m_{m~},m_{m^{\prime}~}\right\}  \left\{  m_{\nu~},m_{\nu^{\prime}~}\right\}
\rho_{m^{\prime}n;\mu\nu^{\prime}~}+\nonumber\\
&  (2\gamma)\underset{m^{\prime},\nu^{\prime}}{%
%TCIMACRO{\tsum }%
%BeginExpansion
{\textstyle\sum}
%EndExpansion
}G_{m_{\mu^{\prime}}-m_{\mu};m_{n^{\prime}}-m_{n}}(\mathbf{R}_{21})\left\{
m_{\mu},m_{\mu^{\prime}}\right\}  \left\{  m_{n},m_{n^{\prime}}\right\}
\rho_{mn^{\prime};\mu^{\prime}\nu}~,
\end{align}
where%
\begin{equation}
\Gamma_{ab~}^{a^{\prime}b^{\prime}}=(2\gamma)\left\{  m_{a~},m_{a^{\prime}%
}\right\}  \left\{  m_{b~},m_{b^{\prime}}\right\}  \delta_{m_{a^{\prime}%
}-m_{a~},m_{b^{\prime}~}-m_{b~}~}.
\end{equation}
Note that some coherence is returned to the ground state as a result of the
"in terms". The derivation of the terms in the right-hand side of Eq.
(\ref{in-terms}) is done by tracing over the field states with one-photon
occupation number, a procedure which has been used in the case of single
multilevel atoms [see for example Ref. \cite{in-terms}]. The first two terms
describe repopulation and recoherence of the ground manifold from the excited
state manifold in a single atom. Cross coupling between atoms is reflected in
the next two terms. Using again the solutions of Eqs. (\ref{AmpEqs}), the
right-hand side of the in-term equations can be expressed in terms of ground
state density matrix elements. A complete system of $16$ linear equations is
thus obtained by adding the in-term to the out-term contributions.

\section{Small Separation ($\mathbf{R}_{21}\ll\lambda$)}

\begin{figure}
[h]
\begin{center}
\includegraphics[
height=2.0557in,
width=3.1298in
]%
{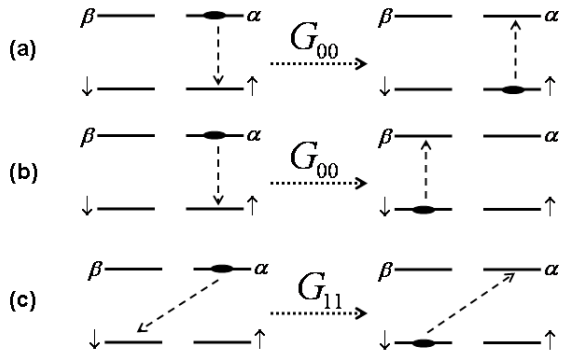}%
\caption{{\footnotesize Illustration of a few possible photon exchanges
between atoms when the interatomic separation is small. Owing to angular
momentum conservation, only diagonal elements of }$G$ {\footnotesize are
present. In (a) and (b) a transfer of excitation involving }$\Delta m=0$
{\footnotesize and, in (c), }$\Delta m=1${\footnotesize , transitions is
shown.}}%
\label{Exchange1}%
\end{center}
\end{figure}

In this limit, owing to angular momentum conservation rules, the propagators
$G_{ij~}$ vanish except for $i=j$. A few photon exchange processes between
atoms are illustrated in Fig. \ref{Exchange1}(c), along with their
accompanying propagators. Notice that due to momentum conservation the
polarization of the emitted photon matches the polarization of the absorbed
photon. Taking as an example the transfer of excitation from atom $1$ in state
$\left\vert \alpha\right\rangle $ (with atom $2$ in state $\left\vert
\downarrow\right\rangle $) to atom $2$ in state $\left\vert \alpha
\right\rangle $ (with atom $1$ in state $\left\vert \downarrow\right\rangle $)
depicted in Fig. \ref{Exchange1}, from Eq. (\ref{in-terms}), one finds that
the propagator associated with the exchange is $G_{11~}$.
%TCIMACRO{\FRAME{fhFU}{3.1298in}{2.0557in}{0pt}%
%{\Qcb{{\footnotesize Illustration of a few possible photon exchanges between
%atoms when the interatomic separation is small. Owing to angular momentum
%conservation, only diagonal elements of }$G$ {\footnotesize are present. In
%(a) and (b) a transfer of excitation involving }$\Delta m=0$
%{\footnotesize and, in (c), }$\Delta m=1${\footnotesize , transitions is
%shown.}}}{\Qlb{Exchange1}}{Fig2.eps}{\special{ language "Scientific Word";
%type "GRAPHIC";  maintain-aspect-ratio TRUE;  display "USEDEF";
%valid_file "F";  width 3.1298in;  height 2.0557in;  depth 0pt;
%original-width 6.6789in;  original-height 4.3785in;  cropleft "0";
%filename 'Fig2.eps';file-properties "XNPEU";}} }%
%BeginExpansion
%EndExpansion

\subsection{Coupled basis}

As in \cite{lehmberg2}, the two-atom system can be described by superpositions
of states that are either symmetrical or antisymmetrical under particle
exchange (Dicke states). The indistinguishability of the particles restricts
the system to the symmetric subspace. The ground state manifold is symmetrized
as follows:%
\begin{align}
\left\vert g_{1~}\right\rangle  &  \equiv\left\vert \uparrow\uparrow
\right\rangle ,\\
\left\vert g_{0}~\right\rangle  &  \equiv\frac{1}{\sqrt{2}}\left(  \left\vert
\uparrow\downarrow\right\rangle +\left\vert \downarrow\uparrow\right\rangle
\right)  ,\nonumber\\
\left\vert g_{-1~}\right\rangle  &  \equiv\left\vert \downarrow\downarrow
\right\rangle ,\nonumber
\end{align}
while ground-excited symmetric states are defined as:%
\begin{align}
\left\vert r_{1~}\right\rangle  &  \equiv\frac{1}{\sqrt{2}}\left(  \left\vert
\alpha\uparrow\right\rangle +\left\vert \uparrow\alpha\right\rangle \right)
,\\
\left\vert r_{-1~}\right\rangle  &  \equiv\frac{1}{\sqrt{2}}\left(  \left\vert
\beta\uparrow\right\rangle +\left\vert \uparrow\beta\right\rangle \right)
,\nonumber\\
\left\vert s_{1~}\right\rangle  &  \equiv\frac{1}{\sqrt{2}}\left(  \left\vert
\alpha\downarrow\right\rangle +\left\vert \downarrow\alpha\right\rangle
\right)  ,\nonumber\\
\left\vert s_{-1~}\right\rangle  &  \equiv\frac{1}{\sqrt{2}}\left(  \left\vert
\beta\downarrow\right\rangle +\left\vert \downarrow\beta\right\rangle \right)
.\nonumber
\end{align}
Rewriting Eqs. (\ref{AmpEqs}) in terms of the new coefficients
$r_{1,-1\text{~}}$and $s_{1,-1\text{~}}$, one finds%
\begin{align}
\overset{\cdot}{s}_{1~}  &  =-\frac{5}{3}\gamma s_{1~}-i\Delta s_{1~}-\frac
{1}{3}\gamma r_{-1~}+i\sqrt{2}\chi g_{-1~},\\
\overset{\cdot}{s}_{-1~}  &  =-\frac{4}{3}\gamma s_{-1~}-i\Delta s_{-1~}%
{},\nonumber\\
\overset{\cdot}{r}_{1~}  &  =-\frac{4}{3}\gamma r_{1~}-i\Delta r_{1~}+i\chi
g_{0~},\nonumber\\
\overset{\cdot}{r}_{-1~}  &  =-\frac{5}{3}\gamma r_{-1~}-i\Delta r_{-1~}%
-\frac{1}{3}\gamma s_{1~},\nonumber
\end{align}
with quasistatic solutions
\begin{align}
s_{1~}  &  =\frac{i\sqrt{2}\chi\left(  \frac{5}{3}\gamma+i\Delta\right)
}{\left(  \frac{5}{3}\gamma+i\Delta\right)  ^{2}-\frac{1}{9}\gamma^{2}}%
g_{-1~},\\
r_{1~}  &  =\frac{i\chi}{i\Delta+\frac{4}{3}\gamma}g_{0~},\nonumber\\
r_{-1~}  &  =O\left[  \left(  \frac{\chi}{\Delta}\right)  ^{2}\right]
,\nonumber\\
s_{-1~}  &  =0.\nonumber
\end{align}
Notice that the symmetric states containing the excited $\beta$ states have
either identically zero or negligibly small amplitude, which substantially
simplifies the calculation.

In the following two subsections, these expressions for the state amplitudes
are used to derive the time evolution of the collective coherence and
population. In the new basis the matrix elements are denoted by $\rho_{ij~}$,
with $i,j=-1,0,1$.

\subsubsection{Coherence decay}

The expression for the collective coherence operator expectation value in the
new basis is given by%
\begin{equation}
\left\langle P_{\downarrow\uparrow~}\right\rangle =\sqrt{2}\left(  \rho
_{10~}+\rho_{0,-1~}\right)  .\label{coh}%
\end{equation}
In the limit $\gamma\ll\Delta$,$~$the following rate equations are obtained
for the out-terms [from Eq. (\ref{out-terms})]%
\begin{align}
\overset{\cdot}{\rho}_{10~}^{out} &  =-\frac{4}{3}\gamma\frac{\chi^{2}}%
{\Delta^{2}}\rho_{10~},\\
\overset{\cdot}{\rho}_{0,-1}^{out} &  =-\frac{14}{3}\gamma\frac{\chi^{2}%
}{\Delta^{2}}\rho_{0,-1~},\nonumber
\end{align}
while the in-terms, obtained from Eq. (\ref{in-terms}), evolve as%
\begin{align}
\overset{\cdot}{\rho}_{10~}^{in} &  =\frac{4}{3}\gamma\frac{\chi^{2}}%
{\Delta^{2}}\rho_{0,-1~},\\
\overset{\cdot}{\rho}_{0,-1~}^{in} &  =\frac{8}{3}\gamma\frac{\chi^{2}}%
{\Delta^{2}}\rho_{0,-1~}.\nonumber
\end{align}
Adding the out-term contribution to the in-terms, and with the notation
$\gamma_{op~}=\gamma\chi^{2}/\Delta^{2}$ (optical pumping rate), the equations
for the density matrix elements relevant for the coherence decay are%
\begin{align}
\overset{\cdot}{\rho}_{10~} &  =-\frac{4}{3}\gamma_{op}\left(  \rho_{10~}%
-\rho_{0,-1~}\right)  ,\label{cohEqs}\\
\overset{\cdot}{\rho}_{0,-1~} &  =-2\gamma_{op}\rho_{0,-1~}.\nonumber
\end{align}
Substituting the solutions of the Eqs. (\ref{cohEqs}) into Eq. (\ref{coh}),
one obtains%
\begin{equation}
\left\langle P_{\downarrow\uparrow~}(t)\right\rangle =\frac{1}{2}%
e^{-2\gamma_{op}t}\left(  -1+3e^{\frac{2}{3}\gamma_{op}t}\right)  .
\end{equation}
This is to be compared with the independent atom coherence decay derived in
Appendix B [Eq. B4]%
\[
\left\langle P_{\downarrow\uparrow~}(t)\right\rangle ^{ind}=e^{-\gamma_{op}t}.
\]%
%TCIMACRO{\FRAME{fhFU}{2.9525in}{2.4215in}{0pt}{\Qcb{{\footnotesize Coupled
%system coherence decay for }$R_{21}\ll\lambda$ {\footnotesize vs. independent
%atoms decoherence.}}}{\Qlb{CohDecayx=0}}{Fig3.eps}%
%width 2.9525in;  height 2.4215in;  depth 0pt;  original-width 3.8726in;
%original-height 3.173in;  cropleft "0";  croptop "1";  cropright "1";
%cropbottom "0";  filename 'Fig3.EPS';file-properties "XNPEU";}} }%
%BeginExpansion
\begin{figure}
[h]
\begin{center}
\includegraphics[
height=2.4215in,
width=2.9525in
]%
{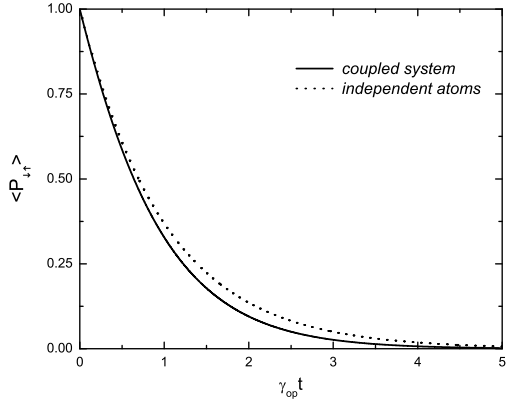}%
\caption{{\footnotesize Coupled system coherence decay for }$R_{21}\ll\lambda$
{\footnotesize vs. independent atoms decoherence.}}%
\label{CohDecayx=0}%
\end{center}
\end{figure}
%EndExpansion
The increase in the coherence decay rate for intermediate times, as shown in
Fig. \ref{CohDecayx=0}, can be understood in terms of the exchange processes
in the uncoupled basis illustrated in Fig. \ref{Exchange1}. For an independent
atom, the state $\left\vert \alpha\right\rangle $ is reached from the
$\left\vert \downarrow\right\rangle $ state through the action of the
classical field, and decays into $\left\vert \downarrow\right\rangle $ and
$\left\vert \uparrow\right\rangle ~$with rates $2(2\gamma/3)$ and
$2(\gamma/3),$ respectively. In the cooperative case, other channels
responsible for coherence generation or decay appear owing to the presence of
the second atom.

An interesting behavior is observed at the initiation stage of the decoherence
process, where the coupled system decoheres at a rate equal to that for
independent atoms. However, this is not a general result, but rather a
consequence of the initial state prepared with equal ground substate
populations. When both atoms start in the same arbitrary state $a\left\vert
\downarrow\right\rangle +b\left\vert \uparrow\right\rangle $ with $a\neq b$
and $a^{2}+b^{2}=1$, the evolution of the coupled system coherence takes the
following form%
\begin{equation}
\left\langle P_{\downarrow\uparrow~}(t)\right\rangle =2abe^{-2\gamma_{op}%
t}\left[  -a^{2}+(1+a^{2})e^{\frac{2}{3}\gamma_{op}t}\right]
,\label{arbitrary}%
\end{equation}
while the independent atom coherence evolves as:%
\begin{equation}
\left\langle P_{\downarrow\uparrow~}(t)\right\rangle ^{ind}=2abe^{-\gamma
_{op}t}.\label{arbitraryindep}%
\end{equation}
Expanding the exponentials in Eqs. (\ref{arbitrary}) and (\ref{arbitraryindep}%
) for small times $\gamma_{op}t\ll1$, we find%
\begin{align}
\left\langle P_{\downarrow\uparrow~}(t)\right\rangle ^{ind} &  \simeq
2ab\left(  1-\gamma_{op}t\right)  ,\nonumber\\
\left\langle P_{\downarrow\uparrow~}(t)\right\rangle  &  \simeq2ab\left[
1-\left(  1+\frac{b^{2}-a^{2}}{3}\right)  \gamma_{op}t\right]  \text{,}%
\end{align}
which shows that the decoherence rate of the coupled system is modified by the
term $(b^{2}-a^{2})/3$. This indicates that, given a population imbalance
between the up and down states, at the initiation stage, the decoherence rate
of the coupled system can be either larger or smaller than the one for the
independent atoms, and vanishes for the balanced case only, when
$a=b=1/\sqrt{2}$.%
%TCIMACRO{\FRAME{fhFU}{3.0035in}{2.5019in}{0pt}{\Qcb{{\footnotesize An
%"imbalanced" superposition state with coefficients }$a,b=1/10,\sqrt{99}%
%/10${\footnotesize ~decays exponentially at a rate }$\gamma_{op}%
%${\footnotesize , under the independent atoms assumption. However, when the
%decay is cooperative, an imbalanced state with }$a<b$%
%{\footnotesize \ \ \ decoheres at a faster rate than one with }$a>b$%
%{\footnotesize . }}}{\Qlb{abstatex=0}}{Fig4.eps}%
%width 3.0035in;  height 2.5019in;  depth 0pt;  original-width 3.8303in;
%original-height 3.186in;  cropleft "0";  croptop "1";  cropright "1";
%cropbottom "0";  filename 'Fig4.EPS';file-properties "XNPEU";}} }%
%BeginExpansion
\begin{figure}
[h]
\begin{center}
\includegraphics[
height=2.5019in,
width=3.0035in
]%
{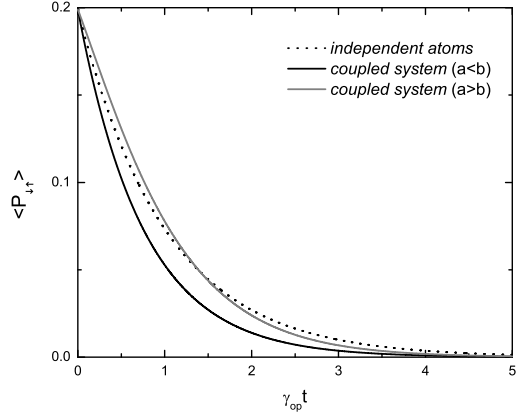}%
\caption{{\footnotesize An "imbalanced" superposition state with coefficients
}$a,b=1/10,\sqrt{99}/10${\footnotesize ~decays exponentially at a rate
}$\gamma_{op}${\footnotesize , under the independent atoms assumption.
However, when the decay is cooperative, an imbalanced state with }%
$a<b${\footnotesize \ \ \ decoheres at a faster rate than one with }%
$a>b${\footnotesize . }}%
\label{abstatex=0}%
\end{center}
\end{figure}
%EndExpansion

\subsubsection{Population Transfer}

In the symmetric basis, the expectation value of the collective population
operator is expressed as%
\begin{equation}
\left\langle P_{\uparrow\uparrow~}\right\rangle =\rho_{11~}+\frac{1}{2}%
\rho_{00~}.\label{pop}%
\end{equation}
Three density matrix elements are coupled to each other: $\rho_{11~}$,
$\rho_{00~}$and $\rho_{-1,-1~}$. The evolution resulting from the classical
field can be obtained [from Eq. (\ref{out-terms})] as%
\begin{align}
\overset{\cdot}{\rho}_{11~}^{out} &  =0,\\
\overset{\cdot}{\rho}_{00}^{out} &  =-\frac{8}{3}\gamma_{op}\rho
_{00~},\nonumber\\
\overset{\cdot}{\rho}_{-1,-1}^{out} &  =-\frac{20}{3}\gamma_{op}\rho
_{-1,-1~},\nonumber
\end{align}
while the in-terms are given by [see Eq. (\ref{in-terms})]%
\begin{align}
\overset{\cdot}{\rho}_{11~}^{in} &  =\frac{4}{3}\gamma_{op}\rho_{00~},\\
\overset{\cdot}{\rho}_{00}^{in} &  =\frac{4}{3}\gamma_{op}\rho_{-1,-1~}%
+\frac{4}{3}\gamma_{op}\rho_{00~},\nonumber\\
\overset{\cdot}{\rho}_{-1,-1}^{in} &  =\frac{16}{3}\gamma_{op}\rho
_{-1,-1~}.\nonumber
\end{align}
Combining the in-terms with out-terms, one finds rate equations%
\begin{align}
\overset{\cdot}{\rho}_{11~} &  =\frac{4}{3}\gamma_{op}\rho_{00~},\\
\overset{\cdot}{\rho}_{00} &  =\frac{4}{3}\gamma_{op}\left(  \rho
_{-1,-1~}-\rho_{00~}\right)  ,\nonumber\\
\overset{\cdot}{\rho}_{-1,-1} &  =-\frac{4}{3}\gamma_{op}\rho_{-1,-1~}%
,\nonumber
\end{align}
that are solved to give%
\begin{equation}
\left\langle P_{\uparrow\uparrow~}(t)\right\rangle =1-\frac{1}{3}e^{-\frac
{4}{3}\gamma_{op}t}\left(  3+2\gamma_{op}t\right)  .
\end{equation}
This is to be compared with the independent atom population evolution%
\begin{equation}
\left\langle P_{\uparrow\uparrow~}(t)\right\rangle ^{ind}=1-e^{-\frac{2}%
{3}\gamma_{op}t}.
\end{equation}
where only one mechanism for populating state $\left\vert \uparrow
\right\rangle $ is present: excitation of state $\left\vert \alpha
\right\rangle $ by the classical field followed by decay at a rate $\gamma/3$
to state $\left\vert \uparrow\right\rangle $.%
%TCIMACRO{\FRAME{fhFU}{3.1237in}{2.5521in}{0pt}{\Qcb{{\footnotesize Cooperative
%optical pumping (from level }$\left\vert \downarrow\right\rangle $
%{\footnotesize to level }$\left\vert \uparrow\right\rangle ${\footnotesize )
%for }$R_{21}\ll\lambda$ {\footnotesize vs. independent atoms optical
%pumping.}}}{\Qlb{opticalpumpingx=0}}{Fig5.eps}%
%width 3.1237in;  height 2.5521in;  depth 0pt;  original-width 3.8571in;
%original-height 3.1453in;  cropleft "0";  croptop "1";  cropright "1";
%cropbottom "0";  filename 'Fig5.EPS';file-properties "XNPEU";}} }%
%BeginExpansion
\begin{figure}
[h]
\begin{center}
\includegraphics[
height=2.5521in,
width=3.1237in
]%
{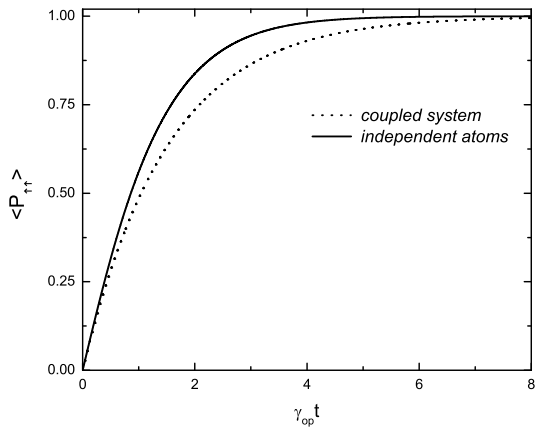}%
\caption{{\footnotesize Cooperative optical pumping (from level }$\left\vert
\downarrow\right\rangle $ {\footnotesize to level }$\left\vert \uparrow
\right\rangle ${\footnotesize ) for }$R_{21}\ll\lambda$ {\footnotesize vs.
independent atoms optical pumping.}}%
\label{opticalpumpingx=0}%
\end{center}
\end{figure}
%EndExpansion

The increase in the population transfer rate, plotted in Fig.
\ref{opticalpumpingx=0} vs. the independent atoms rate, is due to photons
emitted by the second atom that drive the first atom into the $\left\vert
\alpha\right\rangle $ state. A faster excitation of state $\left\vert
\alpha\right\rangle ~$leads to a faster transfer to $\left\vert \uparrow
\right\rangle $ for intermediate times.

\subsection{Mechanisms for coherence generation}

It is interesting to isolate the dynamics of a single atom with the purpose of
identifying the mechanisms that lead to the modification of its radiative
properties due to the presence of a second atom in its vicinity. We analyze
the rate of change of the expectation value of the one atom coherence operator
$\sigma_{-~}^{(1)}=\left\vert \downarrow\right\rangle _{1}\left\langle
\uparrow\right\vert _{1}$. This is expressed in terms of density matrix
elements as $\left\langle \sigma_{-~}^{(1)}\right\rangle =\rho_{\uparrow
\uparrow;\downarrow\uparrow~}+\rho_{\uparrow\downarrow;\downarrow\downarrow~}%
$, and it is found to satisfy the following equation of motion [using Eqs.
(\ref{out-terms}) and (\ref{in-terms})]%
\begin{equation}
\frac{d}{dt}\left\langle \sigma_{-~}^{(1)}(t)\right\rangle =-\gamma
_{op}\left\langle \sigma_{-~}^{(1)}(t)\right\rangle +\frac{1}{3}\gamma
_{op}\left[  \rho_{\downarrow\uparrow;\downarrow\downarrow~}-\rho
_{\uparrow\uparrow;\uparrow\downarrow}\right]  . \label{oneatcoh}%
\end{equation}
The first term in the right-hand side in the above equation simply indicates
the decay of the coherence of the independent atom at the expected rate
$\gamma_{op}$. The second term contains the modification induced by the action
of the neighboring atom. At the moment when the interaction between atoms is
initiated, the density matrix of the coupled system can be factorized and this
term can be written as: $-\frac{1}{3}\gamma_{op}\left\langle \sigma_{z~}%
^{(1)}\right\rangle \left\langle \sigma_{-~}^{(2)}\right\rangle $, \ where
$\sigma_{z~}^{(1)}=\left\vert \uparrow\right\rangle _{1}\left\langle
\uparrow\right\vert _{1}-\left\vert \downarrow\right\rangle _{1}\left\langle
\downarrow\right\vert _{1}$ is the population difference operator for the
first atom. The significance of this term is that $x$ polarization (coherence)
established in the second atom induces $x$ polarization in the first atom
through the vacuum, given a population difference.

We proceed now to analyze the origin of this coupling term by considering two
distinct situations in which atom $2$ is prepared in a superposition
$a\left\vert \downarrow\right\rangle +b\left\vert \uparrow\right\rangle $
exhibiting coherence equal to $ab$, while atom $1$ is prepared either in the
$\left\vert \uparrow\right\rangle $ state or the $\left\vert \downarrow
\right\rangle $ state. In both cases no initial $x$ atomic polarization in the
atom of interest is present. When starting with the population in the
$\left\vert \uparrow\right\rangle $ state, using Eqs. (\ref{out-terms}) and
(\ref{in-terms}), expressions for the density matrix elements present in the
right-hand side of Eq. (\ref{oneatcoh}) can be derived, and an out-term is
found to be responsible with the generation of coherence%
\begin{align}
\frac{d}{dt}\left\langle \sigma_{-~}^{(1)}(t)\right\rangle ^{out}  &
=-\frac{1}{3}\gamma_{op}ab,\label{upEq}\\
\frac{d}{dt}\left\langle \sigma_{-~}^{(1)}(t)\right\rangle ^{in}  &
=0.\nonumber
\end{align}
The process leading to this can be represented as follows%
\begin{equation}
\rho_{\uparrow\uparrow;\uparrow\downarrow~}\overset{\chi}{\longrightarrow}%
\rho_{\uparrow\uparrow;\uparrow\alpha~}\overset{\gamma}{\longrightarrow}%
\rho_{\uparrow\uparrow;\alpha\uparrow~}\overset{\chi}{\longrightarrow}%
\rho_{\uparrow\uparrow;\downarrow\uparrow~}, \label{upprocess}%
\end{equation}
where in the first step, the field induces an excitation in the second atom,
followed by a swap of excitation between atoms through the vacuum
(collision-like effect) and a stimulated emission from the first atom. In the
second case, where the population is initially stored in the $\left\vert
\downarrow\right\rangle $ state, both an in-term and an out-term are present:
\begin{align}
\frac{d}{dt}\left\langle \sigma_{-~}^{(1)}(t)\right\rangle ^{out}  &
=-\frac{1}{3}\gamma_{op}ab,\label{downEq}\\
\frac{d}{dt}\left\langle \sigma_{-~}^{(1)}(t)\right\rangle ^{in}  &  =\frac
{2}{3}\gamma_{op}ab.\nonumber
\end{align}
The coherence generation through the out-term is similar to the process shown
above [Eq. (\ref{upprocess})], while the in-term takes the following path%
\begin{equation}
\rho_{\downarrow\uparrow;\downarrow\downarrow~}\overset{\chi}{\longrightarrow
}\rho_{\downarrow\uparrow;\downarrow\alpha~}\overset{\chi}{\longrightarrow
}\rho_{\alpha\uparrow;\downarrow\alpha~}\overset{\gamma}{\longrightarrow}%
\rho_{\uparrow\uparrow;\downarrow\uparrow~},
\end{equation}
where consecutive excitations for both atoms are followed by spontaneous decay
into a state with coherence in the first atom.

\section{Arbitrary Separation}

\begin{figure}
[h]
\begin{center}
\includegraphics[
height=2.4561in,
width=3.0995in
]%
{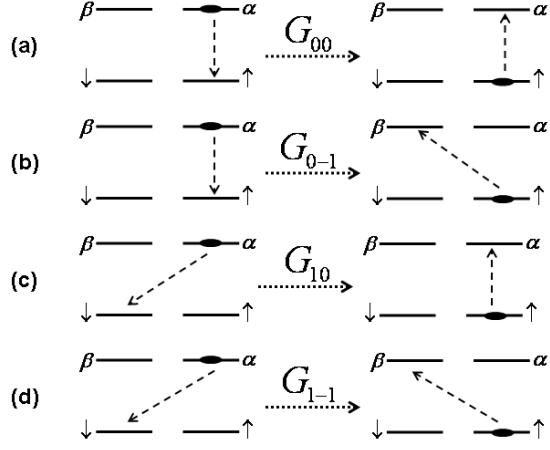}%
\caption{{\footnotesize The transfer of excitation stored in the coupled state
}$\left\vert \alpha\uparrow\right\rangle ~${\footnotesize can take place
through a channel governed by a diagonal propagator }$G_{00}$%
{\footnotesize represented in (a), or through non-diagonal propagators
}$G_{01},$$G_{10}~${\footnotesize and }$G_{1-1}~${\footnotesize as in (b), (c)
and (d).}}%
\label{Exchange2}%
\end{center}
\end{figure}

Simple analytical results are not available in this regime. Coupling through
propagators other than $G_{ii~}$ takes place. The polarization of the emitted
photon doesn't have to match the one of the absorbed photon, a situation which
is illustrated in Fig. \ref{Exchange2}, where the coupling of the two-atom
state $\left\vert \alpha\uparrow\right\rangle $ to other states is shown and
the corresponding elements of the $G$ matrix (including non-diagonal ones)
responsible for the coupling specified.%
%TCIMACRO{\FRAME{fhFU}{3.0995in}{2.4561in}{0pt}{\Qcb{{\footnotesize The
%transfer of excitation stored in the coupled state }$\left\vert \alpha
%\uparrow\right\rangle ~${\footnotesize can take place through a channel
%governed by a diagonal propagator }$G_{00}${\footnotesize represented in (a),
%or through non-diagonal propagators }$G_{01},$$G_{10}~${\footnotesize and
%}$G_{1-1}~${\footnotesize as in (b), (c) and (d).}}}{\Qlb{Exchange2}%
%}{Fig6.eps}{\special{ language "Scientific Word";  type "GRAPHIC";
%width 3.0995in;  height 2.4561in;  depth 0pt;  original-width 6.7646in;
%original-height 5.3532in;  cropleft "0";  croptop "1";  cropright "1";
%cropbottom "0";  filename 'Fig6.eps';file-properties "XNPEU";}} }%
%BeginExpansion
%EndExpansion

\subsection{Numerical Results}

The calculations are now performed in the uncoupled basis. Numerical solutions
of $12$ coupled rate equations give the coherence decay, whereas $13$ rate
equations are solved to obtain the population transfer rate. The output of our
numerical simulations is dependent both on time and on the spherical
coordinates of the second atom $R_{21},\theta~$and $\varphi$. Since the system
has azimuthal symmetry, the interesting cases are obtained by varying $R_{21}$
and $\theta$. Two orientations of the system with respect to the field
propagation are examined: $\mathbf{R}_{21}~\Vert~$$\mathbf{k}_{L}~$($\theta
=0$) and $\mathbf{R}_{21}~\bot~~$$\mathbf{k}_{L}~$($\theta=\pi/2$).

\subsubsection{Coherence Decay}

The $4$ density matrix elements responsible for the coherence [Eq.
(\ref{cohandpop})], are coupled to $8$ more through nondiagonal elements of
$G$. A system of $12$ linear differential equations has to be solved
containing $\left\{  \rho_{\downarrow\downarrow;\downarrow\downarrow\text{ }%
},\rho_{\uparrow\downarrow;\downarrow\downarrow~},\rho_{\downarrow
\uparrow;\downarrow\downarrow~},\rho_{\downarrow\downarrow;\uparrow
\downarrow~},\rho_{\downarrow\downarrow;\downarrow\uparrow~},\rho
_{\uparrow\uparrow;\downarrow\downarrow~},\rho_{\uparrow\downarrow
;\uparrow\downarrow~},\rho_{\uparrow\downarrow;\downarrow\uparrow~}%
,\rho_{\downarrow\uparrow;\downarrow\uparrow~},\rho_{\downarrow\uparrow
;\uparrow\downarrow~},\rho_{\uparrow\uparrow;\uparrow\downarrow~}%
,\rho_{\uparrow\uparrow;\downarrow\uparrow}\right\}  $. The expressions for
all $12$ rate equations are not given here; instead a single one is written to
illustrate the way the coupling among different states comes into play%
\begin{equation}
\overset{\cdot}{\rho}_{\uparrow\downarrow;\downarrow\downarrow~}=\frac
{\gamma_{op}}{3}\left[
\begin{array}
[c]{c}%
-5\rho_{\uparrow\downarrow;\downarrow\downarrow~}-e^{-ik_{L}R_{21}\cos\theta
}\left(
\begin{array}
[c]{c}%
\sqrt{2}G_{10~}(R_{21},\theta,\varphi)\rho_{\downarrow\downarrow
;\downarrow\downarrow~}+2G_{00~}(R_{21},\theta,\varphi)\rho_{\downarrow
\uparrow;\downarrow\downarrow~}\\
+2G_{11~}(R_{21},\theta,\varphi)\rho_{\uparrow\downarrow;\downarrow
\downarrow~}%
\end{array}
\right)  \\
+e^{ik_{L}R_{21}\cos\theta}\left(
\begin{array}
[c]{c}%
2G_{11~}(R_{21},\theta,\varphi)\rho_{\uparrow\downarrow;\downarrow\downarrow
~}+\sqrt{2}G_{01~}(R_{21},\theta,\varphi)\rho_{\uparrow\downarrow
;\downarrow\uparrow~}\\
+\sqrt{2}G_{01~}(R_{21},\theta,\varphi)\rho_{\uparrow\downarrow;\uparrow
\downarrow~}%
\end{array}
\right)
\end{array}
\right]  .
\end{equation}%
%TCIMACRO{\FRAME{fhFU}{2.9265in}{2.4111in}{0pt}{\Qcb{{\footnotesize The
%two-atom coherence for }$k_{L}R_{21}=0.7$ {\footnotesize is plotted as a
%function of time for independent atoms, }$\mathbf{R}_{21}~\Vert~$%
%$\mathbf{k}_{L}~${\footnotesize and }$\mathbf{R}_{21}~\bot~$$\mathbf{k}_{L}%
%${\footnotesize , respectively.\ }}}{\Qlb{IntermediateCohDecay}}%
%{Fig7.eps}{\special{ language "Scientific Word";  type "GRAPHIC";
%width 2.9265in;  height 2.4111in;  depth 0pt;  original-width 3.8571in;
%original-height 3.173in;  cropleft "0";  croptop "1";  cropright "1";
%cropbottom "0";  filename 'Fig7.EPS';file-properties "XNPEU";}} }%
%BeginExpansion
\begin{figure}
[h]
\begin{center}
\includegraphics[
height=2.4111in,
width=2.9265in
]%
{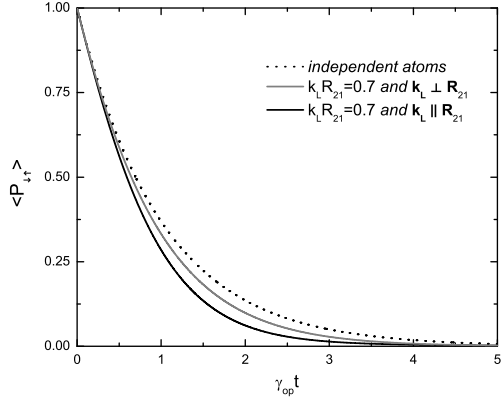}%
\caption{{\footnotesize The two-atom coherence for }$k_{L}R_{21}=0.7$
{\footnotesize is plotted as a function of time for independent atoms,
}$\mathbf{R}_{21}~\Vert~$$\mathbf{k}_{L}~${\footnotesize and }$\mathbf{R}%
_{21}~\bot~$$\mathbf{k}_{L}${\footnotesize , respectively.\ }}%
\label{IntermediateCohDecay}%
\end{center}
\end{figure}
%EndExpansion
\ \ 

\begin{figure}
[h]
\begin{center}
\includegraphics[
height=2.463in,
width=2.8963in
]%
{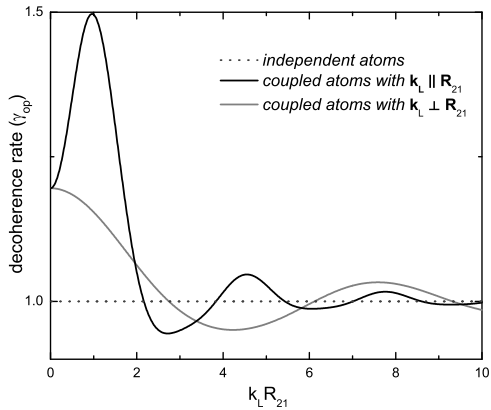}%
\caption{{\footnotesize For a fixed time }$t=1/\gamma_{op}${\footnotesize ,}%
$~${\footnotesize the decoherence rate of the two atoms (in units of }%
$\gamma_{op}${\footnotesize ) is plotted for separations varying from }%
$0~${\footnotesize to }$R_{21}=10/k_{L}${\footnotesize . The dotted line
respresents the independent atom decoherence rate (}$\gamma_{op}%
${\footnotesize ).}}%
\label{ratevardistance}%
\end{center}
\end{figure}

The time evolution of the collective coherence is shown in Fig.
\ref{IntermediateCohDecay} for $k_{L}R_{21}=0.7$. For both $\mathbf{R}%
_{21}~\Vert~$$\mathbf{k}_{L}$ and $\mathbf{R}_{21}~\bot~$$\mathbf{k}_{L}$, the
coherence decays more rapidly than for the independent atom case, and the
decay for $\mathbf{R}_{21}~\Vert~$$\mathbf{k}_{L}$ is faster than that for
$\mathbf{R}_{21}~\bot~$$\mathbf{k}_{L}$. To obtain some idea of the dependence
of the decay rate on interatomic separation, we plot in Fig.
\ref{ratevardistance} the decoherence rate ($\left\langle \overset{\cdot}%
{P}_{\uparrow\uparrow}(t)\right\rangle /\left\langle P_{\uparrow\uparrow
}(t)\right\rangle $) as a function of distance, at a fixed time $t=1/\gamma
_{op}$, for both the $\mathbf{R}_{21}~\Vert~\mathbf{k}_{L}~$and $\mathbf{R}%
_{21~}\bot~\mathbf{k}_{L}$. Even though a numerical solution has been used to
obtain the plot in Fig. \ref{ratevardistance}, an approximate analytical
treatment can provide insight into the qualitative nature of the results. In
particular it can help explain why the parallel case decay rate is larger than
that for closely separated atoms. The coupled equations of motion for the
coherence operators associated with each atom [similar to Eq. (\ref{oneatcoh})
but with the difference that now the coupling coefficients are $R_{21}%
~$dependent] are solved approximately at a fixed time.

It is found that the perpendicular case differs from the close atoms case only
insofar as the coupling between atoms is modulated by the real part of
$G_{00~}(R_{21~},\pi/2,0)$. An analytical expression for a fixed time
$t=1/\gamma_{op}$ gives a decoherence rate of the collective coherence that
varies with the separation as $\gamma_{op}[1+0.21\operatorname{Re}%
(G_{00~}(R_{21},\pi/2,0))]$.

There is, however, a fundamental difference between the perpendicular and
parallel case that is reflected in the system's response. In the perpendicular
case the spatial phase of the laser field does not enter since $\mathbf{k}%
_{L}\cdot\mathbf{R}_{21}=0$. As a result the spin coherence associated with
each atom evolves in an identical fashion. On the other hand, in the parallel
case there is a relative phase difference of $\mathbf{k}_{L}\cdot
\mathbf{R}_{21}$ associated with the laser field as it interacts with atoms at
the two sites. Consequently, the response of the two atoms need no longer be
identical, since the spatial symmetry has been broken by the field. Three
coupling terms are present in the parallel case: $\operatorname{Re}%
(G_{00~}(R_{21},0,0)\cos k_{L}R_{21})$, $\operatorname{Re}(G_{00~}%
(R_{21},0,0)\sin k_{L}R_{21})~$and $\operatorname{Re}(G_{11~}(R_{21},0,0)\sin
k_{L}R_{21})$; the importance of the laser induced spatial phase is evident in
these expressions. Owing to the extra couplings, the two atoms accumulate
different spatial phases, and the collective coherence (obtained as the sum of
individual coherences) shows a spatial modulation that varies as
$\operatorname{Re}(G_{11~}(R_{21},0,0)\sin k_{L}R_{21})$. A full analytical
solution for the variation of the decay rate with the distance for any fixed
time is not available; however, a perturbative treatment for small times
($\gamma_{op}t\ll1$) indicates an increase in the decay rate $2/9$
$\operatorname{Re}(G_{11~}^{2}(R_{21},0,0)\sin^{2}k_{L}R_{21})\gamma_{op}%
^{2}t$.%

%TCIMACRO{\FRAME{fhFU}{2.8963in}{2.463in}{0pt}{\Qcb{{\footnotesize For a fixed
%time }$t=1/\gamma_{op}${\footnotesize ,}$~${\footnotesize the decoherence rate
%of the two atoms (in units of }$\gamma_{op}${\footnotesize ) is plotted for
%separations varying from }$0~${\footnotesize to }$R_{21}=10/k_{L}%
%${\footnotesize . The dotted line respresents the independent atom decoherence
%rate (}$\gamma_{op}${\footnotesize ).}}}{\Qlb{ratevardistance}}{Fig8.eps}%
%width 2.8963in;  height 2.463in;  depth 0pt;  original-width 3.7222in;
%original-height 3.1635in;  cropleft "0";  croptop "1";  cropright "1";
%cropbottom "0";  filename 'Fig8.EPS';file-properties "XNPEU";}} }%
%BeginExpansion
%EndExpansion

\begin{figure}
[h]
\begin{center}
\includegraphics[
height=2.2866in,
width=2.7691in
]%
{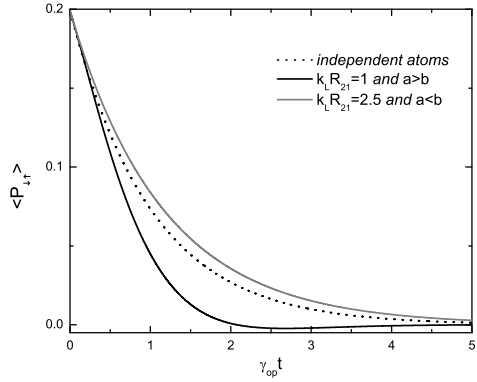}%
\caption{{\footnotesize The decay of the collective coherence of an
"imbalanced" superposition state with coefficients }$1/10~${\footnotesize and
}$\sqrt{99}/10${\footnotesize \ is plotted above. Enhancement of the decay
rate can be obtained when the initial state has more population in the up
state, while the opposite case gives rise to an inhibition of decoherence.}}%
\label{abstatex=1and2.5}%
\end{center}
\end{figure}

Just as in the case of close atoms, we extend our simulations to analyze the
decay of an arbitrary initial state $a\left\vert \downarrow\right\rangle
+b\left\vert \uparrow\right\rangle $. The observed behavior is quite different
here. Even for relatively large separation ($R_{21}=\lambda/2\pi$), a state
with most population in the down state decays much faster than in the
independent atom case, while an inhibition of decoherence is obtained when the
initial state is prepared with more population in the up state [as shown in
Fig. \ref{abstatex=1and2.5} for $\mathbf{R}_{21}~\Vert~$$\mathbf{k}_{L}$ and
$\mathbf{R}_{21}~\bot~$$\mathbf{k}_{L}$].
%TCIMACRO{\FRAME{fhFU}{2.7691in}{2.2866in}{0pt}{\Qcb{{\footnotesize The decay
%of the collective coherence of an "imbalanced" superposition state with
%coefficients }$1/10~${\footnotesize and }$\sqrt{99}/10${\footnotesize \ is
%plotted above. Enhancement of the decay rate can be obtained when the initial
%state has more population in the up state, while the opposite case gives rise
%to an inhibition of decoherence.}}}{\Qlb{abstatex=1and2.5}}{Fig9.eps}%
%width 2.7691in;  height 2.2866in;  depth 0pt;  original-width 3.8138in;
%original-height 3.1453in;  cropleft "0";  croptop "1";  cropright "1";
%cropbottom "0";  filename 'Fig9.EPS';file-properties "XNPEU";}} }%
%BeginExpansion
%EndExpansion

\subsubsection{Optical Pumping}

One more density matrix element ($\rho_{\uparrow\uparrow;\uparrow\uparrow~}$)
is coupled to the $12$ listed before and a system of $13$ rate equations is
solved to obtain the collective population transfer as a function of time.
%TCIMACRO{\FRAME{fhFU}{2.8599in}{2.3091in}{0pt}{\Qcb{{\footnotesize The
%population transfer for }$k_{L}R_{21}=0.7$ {\footnotesize is plotted for
%independent atoms, }$\mathbf{R}_{21}~\Vert~$$\mathbf{k}_{L}~$%
%{\footnotesize and }$\mathbf{R}_{21}~\bot~$$\mathbf{k}_{L}${\footnotesize ,
%respectively.\ }}}{\Qlb{IntermediatePopTransfer}}{Fig10.eps}%
%width 2.8599in;  height 2.3091in;  depth 0pt;  original-width 3.8847in;
%original-height 3.1306in;  cropleft "0";  croptop "1";  cropright "1";
%cropbottom "0";  filename 'Fig10.EPS';file-properties "XNPEU";}} }%
%BeginExpansion
\begin{figure}
[h]
\begin{center}
\includegraphics[
height=2.3091in,
width=2.8599in
]%
{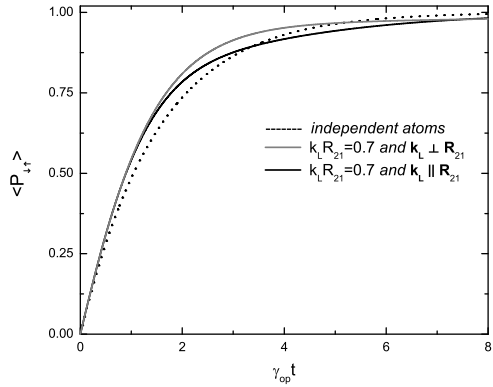}%
\caption{{\footnotesize The population transfer for }$k_{L}R_{21}=0.7$
{\footnotesize is plotted for independent atoms, }$\mathbf{R}_{21}~\Vert
~$$\mathbf{k}_{L}~${\footnotesize and }$\mathbf{R}_{21}~\bot~$$\mathbf{k}_{L}%
${\footnotesize , respectively.\ }}%
\label{IntermediatePopTransfer}%
\end{center}
\end{figure}
%EndExpansion

An interesting behavior is obtained in both the parallel and perpendicular
cases (see Fig. \ref{IntermediatePopTransfer}): the optical pumping rate is
initially larger and afterwards smaller than the one for independent atoms.
The saturation effect at large times is due to the addition of a decay channel
(the $\left\vert \beta\right\rangle ~$state) which provides a way for the
transfer of population from $\left\vert \uparrow\right\rangle $ back to the
initial $\left\vert \downarrow\right\rangle $ state. This accounts for a slow
down at large times where the repopulation from the second atom (resulting in
population in state $\left\vert \beta\right\rangle $) is comparable with the
population produced by the classical field.

\section{Polarization swap}

\begin{figure}
[h]
\begin{center}
\includegraphics[
height=2.4474in,
width=2.9577in
]%
{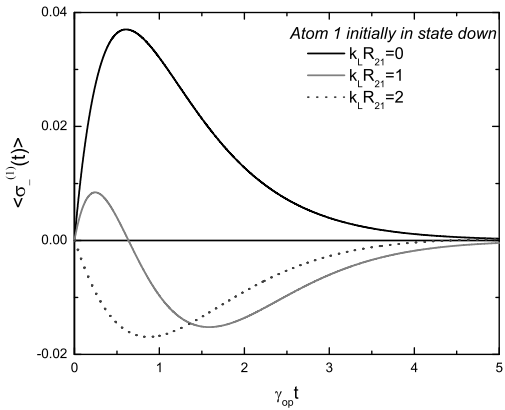}%
\caption{{\footnotesize An }$x~${\footnotesize polarized atom }$2~$%
{\footnotesize induces an }$x~${\footnotesize polarization in an initially
}$z~${\footnotesize polarized atom }$1~${\footnotesize (spin down). The
coherence in atom }$1~${\footnotesize is plotted as a function of time at
}$R_{21}\ll\lambda${\footnotesize , at }$R_{21}=1/k_{L}~${\footnotesize and at
}$R_{21}=2/k_{L}${\footnotesize , for the case when }$\mathbf{R}_{21}~\Vert
~$$\mathbf{k}_{L}$.}%
\label{down}%
\end{center}
\end{figure}

\begin{figure}
[h]
\begin{center}
\includegraphics[
height=2.437in,
width=2.9257in
]%
{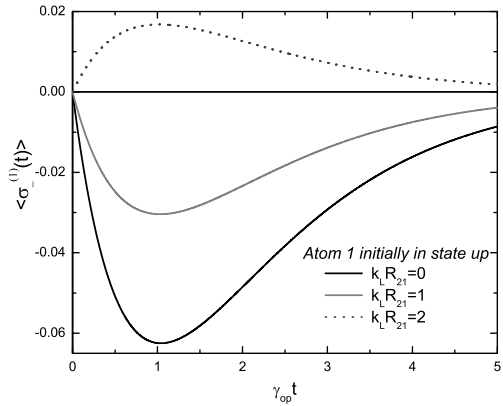}%
\caption{{\footnotesize Atom }$1$ {\footnotesize starts here in state up. The
polarization transfer is plotted as a function of time at }$R_{21}\ll\lambda
${\footnotesize , at }$R_{21}=1/k_{L}~${\footnotesize and }$R_{21}=2/k_{L}%
$,$~${\footnotesize for the case when }$\mathbf{R}_{21}~\Vert~$$\mathbf{k}%
_{L}$.}%
\label{up}%
\end{center}
\end{figure}

As seen in Sec. III B, a partial transfer of coherence from an $x~$polarized
atom to an initially unpolarized atom (in a $z$ state, either up or down) can
be achieved. The two distinct situations discussed there can be extended for
variable interatomic separations. Figures \ref{down} and \ref{up} show the
evolution of the coherence of the first atom as a function of time for
$R_{21}\ll\lambda$ and $R_{21}=\lambda/2\pi$ and $\lambda/\pi$, respectively,
when $\mathbf{R}_{21}~\Vert~$$\mathbf{k}_{L}$. In both cases, owing to the
oscillating nature of the coupling between atoms (as a function of separation)
the sign of the effect produced in the initially $z$ polarized atom depends
dramatically on the distance.%

%TCIMACRO{\FRAME{fhFU}{2.9577in}{2.4474in}{0pt}{\Qcb{{\footnotesize An }%
%$x~${\footnotesize polarized atom }$2~${\footnotesize induces an }%
%$x~${\footnotesize polarization in an initially }$z~${\footnotesize polarized
%atom }$1~${\footnotesize (spin down). The coherence in atom }$1~$%
%{\footnotesize is plotted as a function of time at }$R_{21}\ll\lambda
%${\footnotesize , at }$R_{21}=1/k_{L}~${\footnotesize and at }$R_{21}=2/k_{L}%
%${\footnotesize , for the case when }$\mathbf{R}_{21}~\Vert~$$\mathbf{k}_{L}%
%$.}}{\Qlb{down}}{Fig11.eps}{\special{ language "Scientific Word";
%type "GRAPHIC";  maintain-aspect-ratio TRUE;  display "USEDEF";
%valid_file "F";  width 2.9577in;  height 2.4474in;  depth 0pt;
%original-width 3.8571in;  original-height 3.186in;  cropleft "0";
%filename 'Fig11.EPS';file-properties "XNPEU";}} }%
%BeginExpansion
%EndExpansion
The decay dynamics are still given by Eqs. (\ref{upEq}) and (\ref{downEq}),
with the distinction that the decay parameter, which for close atoms is simply
equal to $\gamma_{op}/3$, is now spatially modulated by $G_{00}(R_{21}%
,0,0)\cos(k_{L}R_{21})$. In the case where the first atom is initially in the
up state, the coherence is driven only by the out-term, which leads to a
simple time behavior, where $\left\langle P_{\uparrow\uparrow}(t)\right\rangle
$ changes sign only due to the spatial oscillation of the coupling term. In
contrast, in the down case, the competition between the in-term and out-term
leads to a change in the sign of $\left\langle P_{\uparrow\uparrow
}(t)\right\rangle $ for intermediate distances.%

%TCIMACRO{\FRAME{fhFU}{2.9257in}{2.437in}{0pt}{\Qcb{{\footnotesize Atom }$1$
%{\footnotesize starts here in state up. The polarization transfer is plotted
%as a function of time at }$R_{21}\ll\lambda${\footnotesize , at }%
%$R_{21}=1/k_{L}~${\footnotesize and }$R_{21}=2/k_{L}$,$~${\footnotesize for
%the case when }$\mathbf{R}_{21}~\Vert~$$\mathbf{k}_{L}$.}}{\Qlb{up}%
%}{Fig12.eps}{\special{ language "Scientific Word";  type "GRAPHIC";
%width 2.9257in;  height 2.437in;  depth 0pt;  original-width 3.8138in;
%original-height 3.173in;  cropleft "0";  croptop "1";  cropright "1";
%cropbottom "0";  filename 'Fig12.EPS';file-properties "XNPEU";}} }%
%BeginExpansion
%EndExpansion

\section{Conclusions\bigskip}

We examined a system of two multilevel atoms coherently driven by a
single-mode classical laser field and coupled to the electromagnetic vacuum.
The decoherence (of a quantum superposition stored in the ground sublevels)
that necessarily accompanies the process of manipulation of the atomic states
has been analyzed in the context of cooperative behavior. Two cases have been
treated, where the atoms are either at the same position or separated by a
distance comparable to the optical wavelength. It has been found (not
surprisingly) that, for the case of close atoms, the "communication" between
atoms leads to an increase in both the decoherence and population transfer
rates. With increasing interatomic separation, in the case of the field
propagating perpendicularly to the line joining the atoms, the decoherence
rate is less than that for close atoms. This is an expected result since, from
the point of view of the classical field the atoms are located at equivalent
positions, and a simple decrease of the interatomic coupling due to the
increasing separation is expected. For certain distances, however, owing to
the oscillating behavior of this coupling, a small effect of decoherence
inhibition is also observed. A more interesting situation arises for the case
when the atoms are aligned parallel to the field propagation direction. The
equivalency of positions does not hold here anymore, and a spatial phase
difference between atoms resulting from the classical field is established.
The coupling through the vacuum is modulated by this spatial phase difference,
and a considerable enhancement in the decay rate is observed at separations of
order $\lambda/2\pi$. These results will be generalized in a future planned
publication to a large ensemble of atoms in a pencil-shaped geometry. In such
a medium, at Fresnel numbers close to unity, the atoms are practically aligned
along the direction of the field; for large optical depths, the phase effect
described above is expected to lead to a substantial increase in the decay of
the collective atomic coherence.

\section{Acknowledgments}

This work is supported by the National Science Foundation under Grant No.
PHY-0244841 and the FOCUS Center grant.

\section{Appendix A: Explicit expressions for propagators}

The expressions for the propagators involve spherical harmonics and Hankel
functions of the first kind, $h_{0}(k_{0}R)$ and $h_{2}(k_{0}R)$ [Ref.
\cite{hao-berman}]:%
\begin{align}
G_{11}(\mathbf{R})  &  =\sqrt{4\pi}h_{0}(k_{0}R)Y_{0,0}(\widehat{\mathbf{R}%
})-\frac{1}{2}\sqrt{\frac{4\pi}{5}}h_{2}(k_{0}R)Y_{2,0}(\widehat{\mathbf{R}%
}),\tag{A1}\\
G_{00}(\mathbf{R})  &  =\sqrt{4\pi}h_{0}(k_{0}R)Y_{0,0}(\widehat{\mathbf{R}%
})+\sqrt{\frac{4\pi}{5}}h_{2}(k_{0}R)Y_{2,0}(\widehat{\mathbf{R}}),\nonumber\\
G_{1,-1}(\mathbf{R})  &  =-\frac{3}{2}\sqrt{\frac{8\pi}{15}}h_{2}%
(k_{0}R)Y_{2,-2}(\widehat{\mathbf{R}}),\nonumber\\
G_{-1,1}(\mathbf{R})  &  =-\frac{3}{2}\sqrt{\frac{8\pi}{15}}h_{2}%
(k_{0}R)Y_{2,2}(\widehat{\mathbf{R}}),\nonumber\\
G_{1,0}(\mathbf{R})  &  =-\frac{3}{2}\sqrt{\frac{4\pi}{15}}h_{2}%
(k_{0}R)Y_{2,-1}(\widehat{\mathbf{R}}),\nonumber\\
G_{-1,0}(\mathbf{R})  &  =-\frac{3}{2}\sqrt{\frac{4\pi}{15}}h_{2}%
(k_{0}R)Y_{2,1}(\widehat{\mathbf{R}}),\nonumber\\
G_{-1,-1}(\mathbf{R})  &  =G_{1,1}(\mathbf{R});\text{ \ }G_{0,-1}%
(\mathbf{R})=-G_{1,0}(\mathbf{R});\text{ \ }G_{0,1}(\mathbf{R})=-G_{-1,0}%
(\mathbf{R}).\nonumber
\end{align}

\section{Appendix B: Derivation of the decoherence and population transfer
rates for a single 4-level atom}

We explicitly derive the equations of motion for the ground state coherence
and populations for a single 4-level atom interacting with a $\sigma_{+}$
polarized field. In the perturbative limit where a maximum of one excitation
is allowed, a basis set in the Hilbert space of atom and radiation field is
comprised of states $\left\vert \downarrow\right\rangle \left\vert
0\right\rangle $, $\left\vert \uparrow\right\rangle \left\vert 0\right\rangle
\,$, $\left\vert \alpha\right\rangle \left\vert 0\right\rangle $, $\left\vert
\beta\right\rangle \left\vert 0\right\rangle $, $\left\vert \downarrow
\right\rangle \left\vert \mathbf{k},\lambda\right\rangle $ and $\left\vert
\uparrow\right\rangle \left\vert \mathbf{k},\lambda\right\rangle $. The first
ket denotes the state of the atom while, the second one describes a vacuum
field or a one photon state with wave number $\mathbf{k}$ and polarization
$\lambda$. The state amplitudes obey the following equations of motion:
\begin{align}
\overset{\cdot}{b}_{\alpha;0~}  &  =-\gamma b_{\alpha;0~}-i\Delta
b_{\alpha;0~}+i\chi b_{\downarrow0~},\tag{B1}\\
\overset{\cdot}{b}_{\beta;0~}  &  =-\gamma b_{\beta;0~}-i\Delta b_{\beta
;0~},\nonumber\\
\overset{\cdot}{b}_{\downarrow;0~}  &  =i\chi b_{\alpha;0~},\nonumber\\
\overset{\cdot}{b}_{\uparrow;0~}  &  =0,\nonumber\\
\overset{\cdot}{b}_{\downarrow;\mathbf{k},\lambda~}  &  =ig_{\mathbf{k}%
,\lambda~}^{\downarrow\alpha}b_{\alpha;0~},\nonumber\\
\overset{\cdot}{b}_{\uparrow;\mathbf{k},\lambda~}  &  =ig_{\mathbf{k}%
,\lambda~}^{\uparrow\beta}b_{\beta;0~}.\nonumber
\end{align}
Using a master equation approach one can write density matrix equations of
motion for the ground state sublevels as:%
\begin{align}
\overset{\cdot}{\rho}_{\uparrow\uparrow~}  &  =2\left(  \frac{2\gamma}%
{3}\right)  \rho_{\beta\beta~}+2\left(  \frac{\gamma}{3}\right)  \rho
_{\alpha\alpha~},\tag{B2}\\
\overset{\cdot}{\rho}_{\downarrow\uparrow~}  &  =-2\left(  \frac{2\gamma}%
{3}\right)  \rho_{\alpha\beta~}+i\chi\rho_{\alpha\uparrow~},\nonumber\\
\overset{\cdot}{\rho}_{\downarrow\downarrow~}  &  =2\left(  \frac{\gamma}%
{3}\right)  \rho_{\beta\beta~}+2\left(  \frac{2\gamma}{3}\right)  \rho
_{\alpha\alpha~}+i\chi\left(  \rho_{\downarrow\alpha}-\rho_{\alpha\downarrow
}\right)  .\nonumber
\end{align}
The first observation that we make here is that the terms in the right hand
side of the above equations, that are due to the field (out-terms) and to the
coupling to the vacuum (in-terms) can be derived separately. In other words,
the presence of dissipation can be neglected when writing equations describing
the driving effect of the field and later added to the equations
phenomenologically. The derivation of rate equations can now be carried out by
writing equations for the ground-excited coherences and excited state
populations and coherences and adiabatically eliminating them. However, this
is an unnecessary complication; the second observation that we make provides
an easier way of doing this. Only the first $4$ equations in Eqs. (B1) have to
be solved and the excited amplitudes can be written in terms of ground state
amplitudes. Next, the density matrix elements can be written simply as if the
state of the system were a pure state (as products of amplitudes) and replaced
in Eqs. (2) to obtain rate equations. The separation of in-terms from
out-terms in these equations insures the validity of this approach. Following
this recipe, it is found that (in the limit $\Delta\gg\gamma$)%
\begin{align}
\rho_{\alpha\beta~}  &  =b_{\alpha;0~}b_{\beta;0~}^{\ast}=0,\tag{B3}\\
\rho_{\alpha\alpha~}  &  =b_{\alpha;0~}b_{\alpha;0~}^{\ast}\simeq\frac
{\chi^{2}}{\Delta^{2}}b_{\downarrow0~}b_{\downarrow0~}^{\ast}=\frac{\chi^{2}%
}{\Delta^{2}}\rho_{\downarrow\downarrow~},\nonumber\\
\rho_{\beta\beta~}  &  =b_{\beta;0~}b_{\beta;0~}^{\ast}=0,\nonumber\\
\rho_{\alpha\uparrow~}  &  =b_{\alpha;0~}b_{\uparrow;0~}^{\ast}\simeq
i\gamma\frac{\chi}{\Delta^{2}}b_{\downarrow0~}b_{\uparrow0~}^{\ast}+\frac
{\chi}{\Delta}b_{\downarrow0~}b_{\uparrow0~}^{\ast}=i\gamma\frac{\chi}%
{\Delta^{2}}\rho_{\downarrow\uparrow~}+\frac{\chi}{\Delta}\rho_{\downarrow
\uparrow~},\nonumber\\
\rho_{\downarrow\alpha~}  &  =b_{\downarrow;0~}b_{\alpha;0~}^{\ast}%
=i\gamma\frac{\chi}{\Delta^{2}}b_{\downarrow0~}b_{\downarrow0~}^{\ast}%
+\frac{\chi}{\Delta}b_{\downarrow0~}b_{\downarrow0~}^{\ast}=i\gamma\frac{\chi
}{\Delta^{2}}\rho_{\downarrow\downarrow~}+\frac{\chi}{\Delta}\rho
_{\downarrow\downarrow~},\nonumber\\
\rho_{\alpha\downarrow~}  &  =\rho_{\downarrow\alpha~}^{\ast}.\nonumber
\end{align}
Replacing these expressions into Eqs. (B2), rate equations are obtained in a
final form:%
\begin{align}
\overset{\cdot}{\rho}_{\uparrow\uparrow~}  &  =\frac{2}{3}\gamma_{op}%
\rho_{\downarrow\downarrow~},\tag{B4}\\
\overset{\cdot}{\rho}_{\downarrow\uparrow~}  &  =-\gamma_{op}\rho
_{\downarrow\uparrow~}+i\frac{\chi^{2}}{\Delta}\rho_{\downarrow\uparrow
~},\nonumber\\
\overset{\cdot}{\rho}_{\downarrow\downarrow~}  &  =\frac{4}{3}\gamma_{op}%
\rho_{\downarrow\downarrow~}-2\gamma_{op}\rho_{\downarrow\downarrow~}%
=-\frac{2}{3}\gamma_{op}\rho_{\downarrow\downarrow~}.\nonumber
\end{align}
The decoherence of an initial state $\frac{1}{\sqrt{2}}\left(  \left\vert
\downarrow\right\rangle +\left\vert \uparrow\right\rangle \right)  $ can now
be calculated as described by the evolution of $\rho_{\downarrow\uparrow~}%
$(neglecting the phase associated with the AC Stark shift)
\begin{equation}
\rho_{\downarrow\uparrow~}(t)=\frac{1}{2}e^{-\gamma_{op}t}, \tag{B5}%
\end{equation}
while the population transfer from state $\left\vert \downarrow\right\rangle $
to state $\left\vert \uparrow\right\rangle $ is given by%
\begin{equation}
\rho_{\uparrow\uparrow~}(t)=1-e^{-\frac{2}{3}\gamma_{op}t}. \tag{B6}%
\end{equation}


\begin{thebibliography}{99}
\bibitem {footnote}It should be noted that the qualitative results for this
particular case cannot be generalized to other level scheme or field
polarization. For example, a calculation on two close atoms (not presented
here) that uses $x$ instead of $\sigma_{+}$ polarized driving fields and the
same $J=1/2\rightarrow J^{\prime}=1/2$ transition, indicates that the
collective decoherence rate is not different from the independent atom
coherence decay rate.
\end{thebibliography}
\end{document}